\documentclass[10pt]{article}
 \usepackage{graphicx}
\usepackage{amsfonts}
\usepackage{amssymb}
\usepackage{amsmath}

\begin{document}
\title{Polarization mode interaction equations in optical fibers with Kerr effect}%
\author{S. B. Leble, B. Reichel\\ \small
 Technical University of Gda\'{n}sk\\\small
   ul. G.Narutowicza, 11  80-952, Gda\'{n}sk-Wrzeszcz, Poland\\}
 \maketitle
\begin{abstract}
We derive coupled nonlinear Schr\"{o}dinger equation (CNLSE) for
arbitrary polarized light propagation in a single-mode fiber. We
introduce a basis of transverse eigen modes with the appropriate
projecting hence solutions depend on the waveguide geometry.
Considering a weak nonlinearity which is connected with Kerr
effect, we give explicit expressions for nonlinear constants via
integrals of Bessel functions. We compare numerical results for
the nonlinear constant extracted from experimental observations of
a soliton for the nonlinear Schr\"{o}dinger equation (NLSE)
(single-mode one). The method of  projecting we use allows a
direct generalization to multi-mode fiber case.
\end{abstract}

\section{Introduction}
There are lot of publications devoted to the propagation and
interaction of polarized electromagnetic wave pulses
 in optical fibers (see the book \cite{Porsezian:OpticalSolitons} ). Almost most of them exploit the results of \cite{refMenyuk},
that is going up to \cite{HasegawaKodama1981}. It is claimed that
one can consider the fiber as made of isotropic material and the
birefringence is originated from the third order nonlinearity
(Kerr effect \cite{agrawal:book:NonFibOpt}). The result of the
derivation is achieved by means of averaging across the fiber
section and gives the following evolution along z axis.
\begin{subequations}\label{NSo}
\begin{eqnarray}
iX^{+}_{z}-ik'^{+}_{N}X^{+}_{t}+\frac{k''}{2}X^{+}_{tt}+\left(\gamma|X^{+}|^{2}+\eta|X^{-}|^{2}\right)X^{+}+\dots&=&0,\\
iX^{-}_{z}-ik'^{-}_{N}X^{-}_{t}+\frac{k''}{2}X^{-}_{tt}+\left(\gamma|X^{-}|^{2}+\eta|X^{+}|^{2}\right)X^{-}+\dots&=&0,
\end{eqnarray}
\end{subequations}
where $X^{+}, X^{-}$ are the envelopes components of electric
fields (polarizations), $k''$ is the dispersion constants
respectively, $\gamma$ correspond to SPM (self phase modulation)
and $\eta$ corresponds to XPM (cross phase modulation). The
computation of the $\gamma/\eta$ relation (ratio) for Kerr medium
which is generally elliptically birefringent, depends on
birefringent ellipse axis choice \cite{refMenyuk, refMenyuk1,
refMenyuk2, refBlowStabNLS} and have value  2/3 (linear case), 2
(circular case), generally (for example $\gamma/\eta=2$) we can
say that the XPM is twice as effective as SPM .
$V^{\pm}_{gN}=1/k'^{\pm}_{N}$ is the nonlinear group velocity of
polarization components. Here we accept that the origin of a
birefringence comes from nonlinear effects, but it can be also
descent from random defects of fiber or special structure of
waveguide (polarization maintaining fibers).

The averaging procedure looks reasonable from a physical scope but
in many cases leads to significant deviations from experiments
\cite{leble:book:NonWaves}. The transition from a
three-dimensional to one-dimensional picture by the averaging is
quite impossible in the case of multi-mode field: it leads to the
only equation while the modes should be described by independent
variables.

In this paper we base on a projecting procedure to the mode
subspaces in a functional space of a multi-mode field
\cite{leble:book:NonWaves}. In the nonlinear theory it leads to
the important difference between results for nonlinear constants
obtained by the projecting and averaging procedures already in the
case of one-mode fiber.

 The general plan of the paper is
following. In the section \ref{SEC:Basic} we briefly overview
(fix) the notation that are chosen maximally close to standard
books \cite{agrawal:book:NonFibOpt}.
 In the next section \ref{SEC:GenSolution} we show how to build the general
representation of the overall electromagnetic field as the mode
superposition. We also define the transverse orthogonal eigenmodes
together with the projecting procedure by the appropriate scalar
product. The section \ref{SEC:Numerical} contains numerical
results for a single
 mode waveguide and the evaluation of nonlinear coefficients for the NLS
 equation.

\section{Basic equations}\label{SEC:Basic}
 We   describe
polarization modes interaction in the cylindrical optical fibers.
We start from the Maxwell  electromagnetic field equations
\begin{subequations} \label{E:MaxwellEquation}
\begin{eqnarray}
\label{Eq:MaxdivB}
\nabla\cdot\mathbf B&=&0,\\
\label{Eq:MaxdivD}
\nabla\cdot\mathbf D&=&0,\\
\label{Eq:MaxrotE}
\nabla\times\mathbf E&=&-\frac{\partial{\mathbf B}}{\partial t},\\
\label{Eq:MaxrotH}
 \nabla\times\mathbf H&=&\frac{\partial{\mathbf
D}}{\partial t},
\end{eqnarray}
\end{subequations}
in the system in the cylindrical polar coordinate ($r$, $\varphi$,
$z$)\\
 and materials equations
\begin{subequations} \label{E:MaterialEquation}
\begin{eqnarray}
\mathbf H&=&\frac{1}{\mu_{0}}\mathbf B,\\
\mathbf D &=& \varepsilon_{0} \mathbf E+\mathbf{P}.
\end{eqnarray}
\end{subequations}
When one study boundary conditions the polarization vector
$\mathbf{P}$ is considered as a linear function of $\mathbf{E}$,
we take the simplest form for the isotropic medium
\begin{equation}
   \mathbf{P}=\varepsilon_{0}\chi_{linear}\mathbf E
\end{equation}
 and a wave equation for the electric field in fiber is
\begin{equation} \label{E:WaveEquation}
\Delta\mathbf{E}-\mu_{0}\varepsilon_{0}\varepsilon\frac{\partial^{2}\mathbf{E}}{\partial{t}^{2}}=0.
\end{equation}
Boundary conditions for our waveguide are
\begin{subequations}  \label{E:BoundaryCond}
\begin{eqnarray}
D_{r1}-D_{r2}&=&0,\\
B_{r1}-B_{r2}&=&0,\\
\label{sE:ga1}
\mathbf{n}\times(\mathbf{E}_{2}-\mathbf{E}_{1})&=&0,\\
\label{sE:ga2}
\mathbf{n}\times(\mathbf{H}_{2}-\mathbf{H}_{1})&=&0.
\end{eqnarray}
\end{subequations}
Conditions \eqref{sE:ga1} for electric field yield
\begin{subequations} \label{E:BoundaryPHIRALL}
\begin{eqnarray}
\varepsilon_{1}E_{r}(r_{0+})&=&\varepsilon_{2}E_{r}(r_{0-}),\\
E_{\varphi}(r_{0+})&=&E_{\varphi}(r_{0-}),\\
E_{z}(r_{0+},\varphi,z)&=&E_{z}(r_{0-},\varphi,z)\text{\qquad
($r_{0}$ - waveguide radius)}.
\end{eqnarray}
\end{subequations}

 Wave number $k$ must be the same inside and outside a
waveguide. To perform boundary conditions, we defined two
parameters $\alpha$ and $\beta$
\begin{subequations} \label{E:BounduaryKvec}
\begin{eqnarray}
\alpha^{2}&=&\omega^{2}\varepsilon_{0}\mu_{0}\varepsilon_{1}-k^{2}\text{,\qquad $r\leq r_{0}$},\\
\beta^{2}&=&k^{2}-\omega^{2}\varepsilon_{0}\mu_{0}\varepsilon_{2}
\text{,\qquad $r>r_{0}$},
\end{eqnarray}
\end{subequations}
where $\omega$ is the frequency of a light wave.

Now if we use solution inside and outside then waveguide for
linear polarization and all of boundary conditions we get equation
(known as Hondros-Debye equation) for the eigenvalues
$\alpha_{ln}$ (see equations \ref{E:BounduaryKvec}). From this
equation (which is well known in linear theory of waveguides
\cite{agrawal:book:NonFibOpt}) we can numerically evaluate
eigenvalues. The linearized ME defined the basis of eigenfunctions
that are
\begin{equation}
J_{l}(\alpha_{ln}r)e^{il\varphi}
\end{equation}
where $l=0,\pm 1,\pm 2,\dots$ and $n= 1,2,\dots$ where $n$ is
numbering  following eigenvalues (following solutions for fixed
$l$) and $\alpha_{ln}$ is connected with eigenvalues $k_{ln}$.

In general the polarization vector should be written as
\begin{equation} \label{E:anizpolnonlin}
\mathbf P=\varepsilon_{0}\left(\chi^{(1)}\mathbf
E+\chi^{(2)}\vdots \mathbf E\mathbf E+\chi^{(3)}\vdots\mathbf
E\mathbf E\mathbf E+\cdots\right),
\end{equation}
where $\chi^{(1)}$ is linear dielectric susceptibility and
corresponds to the  refraction of light. In a case of the second
order dielectric susceptibility $\chi^{(2)}$, we could omit it
because it is equal zero in materials construct with symmetrical
molecule. From higher order dielectric susceptibility we save only
third order susceptibility because rest of orders are negligible.
The third order susceptibility is responsible for nonlinear
refraction of light, self phase modulation (SPM) and cross phase
modulation (XPM).

If we take into consideration that for impulses longer then 0.1ps
one can treat a response of a medium as instantaneous and we can
write
\begin{equation} \label{E:Polarization3_slow}
\mathrm{\mathbf{P}_{NL}}(t)=\varepsilon_{0}\chi^{(3)}(t,t,t)\vdots\mathbf
E(t)\mathbf E(t)\mathbf E(t).
\end{equation}
The third order dielectric susceptibility $\chi^{(3)}$ for
isotropic media is discussed in the papers
\cite{agrawal:book:NonFibOpt, Terhune}. Basing on it we write
\begin{equation}\label{lab:war1}
\chi_{ijkl}=\chi_{xxxx}\delta_{ij}\delta_{kl}+\chi_{xyxy}\delta_{ik}\delta_{jl}+\chi_{xyyx}\delta_{il}\delta_{kj},
\end{equation}
\begin{equation}\label{lab:war2}
\chi_{xxxx}=\chi_{yyyy}=\chi_{zzzz}=\chi_{xxyy}+\chi_{xyxy}+\chi_{xyyx},
\end{equation}
\begin{equation}\label{lab:war3}
\chi_{xxyy}\simeq\chi_{xyxy}\simeq\chi_{xyyx},
\end{equation}
all components of electric field are in  standard form
\begin{equation}
 E_{i}=\frac{1}{2}A_{i}e^{i\omega t}+c.c.,
\end{equation}
inserting this relation into equation
\eqref{E:Polarization3_slow}, we get nonlinear polarization as
(non-resonant terms are removed)
\begin{equation} \label{Pol_unknow}
P_{i}=\frac{1}{8}\chi_{xxxx}\varepsilon_{0}\sum_{j}\left(2A_{i}|A_{j}|^{2}+A_{j}^{2}\overline{A}_{i}\right)e^{i\omega
t}+c.c.,
\end{equation}
where $i,j=x,y,z$

For example, for $z$ component we have
\begin{equation}
P_{z}=\frac{3}{8}\chi_{xxxx}\varepsilon_{0}\left\{\left[|A_{z}|^{2}+\frac{2}{3}\left(|A_{x}|^{2}+|A_{y}|^{2}\right)\right]A_{z}+\frac{1}{3}\overline{A}_{z}\left(A_{x}^{2}+A_{y}^{2}\right)\right\}e^{i\omega
t}+c.c..
\end{equation}
We have same equation as in paper \cite{agrawal:book:NonFibOpt,
refMenyuk}. Next we put it into the Maxwell equations (We
introduce nonlinearity into the Maxwell equations in the form of
the Kerr effect \cite{agrawal:book:NonFibOpt}, with assumption of
small nonlinearity.)

Let us rewrite the wave equation system as
\begin{equation}\label{E:NonLinWave}
\square E_{i}=-\mu_{0}\frac{\partial^{2}}{\partial t^{2}}P_{i},
\end{equation}
where $\square$ is defined by
\begin{equation}\label{EQ:operator1}
\square =
\mu_{0}\varepsilon_{0}\varepsilon\frac{\partial^{2}}{\partial
t^{2}} -  \triangle.
\end{equation}

\section{General solution, main results}\label{SEC:GenSolution}
  We write solution for electromagnetic field
 with amplitude depend on time and propagation coordinate in form
 \cite{leble:book:NonWaves}:
\begin{subequations} \label{E:WspAL}
\begin{multline}
E_{z}(r,\varphi,z,t)=\frac{1}{2}\sum_{p,l,n}\left[\mathcal{A}_{ln}^{p}(z,t)J_{l}(\alpha_{nl}
r)e^{il\varphi}e^{i(\omega t-kz)}+c.c.\right],\\
\end{multline}
\begin{multline}
E_{r}(r,\varphi,z,t)=\\-\frac{1}{2}\sum_{p,l,n}\left\{\frac{i}{\alpha_{nl}^{2}}\left[\tilde{\mathcal{B}}_{ln}^{p}(z,t)\frac{il\omega}{r}J_{l}(\alpha_{nl}
r)+\tilde{\mathcal{C}}^{p}_{ln}(z,t)k\partial_{r}J_{l}(\alpha_{nl}
r)\right]e^{il\varphi}e^{i(\omega t-kz)}+c.c.\right\},
\end{multline}
\begin{multline}
E_{\varphi}(r,\varphi,z,t)=\\\frac{1}{2}\sum_{p,l,n}\left\{\frac{i}{\alpha_{nl}^{2}}\left[\tilde{\mathcal{D}}^{p}_{
ln}(z,t)\omega\partial_{r} J_{l}(\alpha_{nl}
r)-\tilde{\mathcal{E}}^{p}_{ln}(z,t)\frac{ilk}{r}J_{l}(\alpha_{nl}
r)\right]e^{il\varphi}e^{i(\omega t-kz)}+c.c.\right\},
\end{multline}
\begin{multline}
B_{z}(r,\varphi,z,t)=\frac{1}{2}\sum_{p,l,n}\left[\mathcal{F}^{p}_{ln}(z,t)J_{l}(\alpha_{nl}
r)e^{il\varphi}e^{i(\omega t-kz)}+c.c.\right],\\
\end{multline}
\begin{multline}
B_{r}(r,\varphi,z,t)= \\
\frac{1}{2}\sum_{p,l,n}\left\{\frac{i}{\alpha_{nl}^{2}}\left[\tilde{\mathcal{G}}^{p}_{ln}(z,t)\frac{il\omega
\mu_{o}\varepsilon_{0}\varepsilon}{r}J_{l}(\alpha_{nl}
r)-\tilde{\mathcal{H}}^{p}_{ln}(z,t)k\partial_{r}J_{l}(\alpha_{nl}
r)\right]e^{il\varphi}e^{i(\omega t-kz)}+c.c.\right\},
\end{multline}
\begin{multline}
B_{\varphi}(r,\varphi,z,t)=\\-\frac{1}{2}\sum_{p,l,n}\left\{\frac{i}{\alpha_{nl}^{2}}\left[\tilde{\mathcal{P}}^{p}_{
ln}(z,t)\omega\mu_{o}\varepsilon_{0}\varepsilon
\partial_{r}J_{l}(\alpha_{nl}
r)+\tilde{\mathcal{S}}^{p}_{ln}(z,t)\frac{ilk}{r}J_{l}(\alpha_{nl}
r)\right]e^{il\varphi}e^{i(\omega t-kz)}+c.c.\right\}.
\end{multline}
\end{subequations}
Here $p$ numbering two orthogonal polarization and have values
''$+$'' and ''$-$''. Coefficients with tilde includes all
constants to simplify notation.

 Inserting these solutions into the Maxwell equations yields
\begin{subequations} \label{E:VarEqMax_a}
\begin{eqnarray}
-\mathcal{B}^{p}_{ln}+il\mathcal{D}^{p}_{ln}=0,\\
\mathcal{C}^{p}_{ln}\alpha^{2}_{ln}+\partial_{z}\mathcal{A}^{p}_{ln}=0,\label{SS:VdivE}\\
-\mathcal{C}^{p}_{ln}l^{2}-\mathcal{E}^{p}_{ln}il=0,\\
il\mathcal{A}^{p}_{ln}+\partial_{z}\mathcal{E}^{p}_{ln}-\partial_{t}\mathcal{H}^{p}_{ln}=0,\\
-\partial_{z}\mathcal{D}^{p}_{ln}+\partial_{t}\mathcal{G}^{p}_{ln}=0,\\
-\partial_{z}\mathcal{C}^{p}_{ln}-\mathcal{A}^{p}_{ln}-\partial_{t}\mathcal{P}^{p}_{ln}=\label{SS:CzAPt}0,\\
-\partial_{z}\mathcal{B}^{p}_{ln}-\partial_{t}\mathcal{S}^{p}_{ln}=0,\\
-\mathcal{D}^{p}_{ln}\alpha^{2}_{ln}+\partial_{t}\mathcal{F}^{p}_{ln}=0,\\
\mathcal{D}^{p}_{ln}l^{2}+il\mathcal{B}^{p}_{ln}=0,\\
-\mathcal{E}^{p}_{ln}+il\mathcal{G}^{p}_{ln}=0,\\
\mathcal{G}^{p}_{ln}-il\mathcal{P}^{p}_{ln}=0,\\
\mathcal{H}^{p}_{ln}\alpha^{2}_{ln}+\partial_{z}\mathcal{F}^{p}_{ln}=0,\\
-\mathcal{H}^{p}_{ln}l^{2}-\mathcal{S}^{p}_{ln}il=0,\\
\mathcal{F}^{p}_{ln}il+\partial_{z}\mathcal{S}^{p}_{ln}+\mu_{0}\varepsilon_{0}\varepsilon\partial_{t}\mathcal{B}^{p}_{ln}=0,\\
\partial_{z}\mathcal{P}^{p}_{ln}+\mu_{0}\varepsilon_{0}\varepsilon\partial_{t}\mathcal{C}^{p}_{ln}=0,\\
\partial_{z}\mathcal{G}^{p}_{ln}+\partial_{t}\mathcal{E}^{p}_{ln}=0,\\
-\partial_{z}\mathcal{H}^{p}_{ln}-\mathcal{F}^{p}_{ln}-\mu_{0}\varepsilon_{0}\varepsilon\partial_{t}\mathcal{D}^{p}_{ln}=0,\\
\alpha^{2}_{ln}\mathcal{P}^{p}_{ln}-\mu_{0}\varepsilon_{0}\varepsilon\partial_{t}\mathcal{A}^{p}_{ln}=0\label{SS:VrotBR},\\
-\mathcal{P}^{p}_{ln}l^{2}-\mathcal{G}^{p}_{ln}il=0,\\
-\mathcal{S}^{p}_{ln}+\mathcal{H}^{p}_{ln}il=0.
\end{eqnarray}
\end{subequations}

Using \eqref{SS:VdivE}, \eqref{SS:VrotBR} and  \eqref{SS:CzAPt} we
can verify that the amplitudes $\mathcal{A}^{p}_{ln}$ satisfy
equation (similarly we can get equation for
$\mathcal{F}^{p}_{ln}$)
\begin{equation} \label{E:VEM}
\partial_{zz}\mathcal{A}^{p}_{ln}-\mu_{0}\varepsilon_{0}\varepsilon
\partial_{tt}\mathcal{A}^{p}_{ln}=\alpha^{2}_{ln}\mathcal{A}^{p}_{ln}
\end{equation}

It can be proved that the Bessel functions satisfy  orthogonality
in form \cite{SpecialFunctions}
\begin{equation}\label{E:BesselOrhtoNormal}
\begin{split}
\int\limits_{0}^{r_{0}}rJ_{l}(\alpha_{ln^{'}}r)J_{l}(\alpha_{ln}
r)dr=\frac{r_{0}^{2}}{2}\left[J_{l}^{2}(\alpha_{ln}
r_{0})-J_{l-1}(\alpha_{ln} r_{0})J_{l+1}(\alpha_{ln}
r_{0})\right]\delta_{n^{'}n}=N_{nl}\delta_{n^{'}n},
\end{split}
\end{equation}
taking into account boundary condition for optical waveguide.

Let us now exploit orthogonal relation, we can show that the
equation for $z$ coordinate are \cite{leble:book:NonWaves}
\begin{equation}\label{EQ:Equ_z_nonlin}
\left(\square_{z}+\alpha^{2}_{ln}\right)\mathcal{A}^{p}_{ln}=\frac{2\varepsilon_{0}\mu_{0}}{\pi
N_{nl}}\int\limits_{0}^{r_{0}}\int\limits_{0}^{2\pi}
rJ_{l}(\alpha_{ln} r)e^{-il\varphi}\frac{\partial^2}{\partial
t^2}\sum_{klm}\chi_{zklm}E_{k}E_{l}E_{m}d\varphi dr,
\end{equation}
where $\square_{z}$ is defined by
\begin{equation}\label{EQ:operator1z}
\square_{z} =
\mu_{0}\varepsilon_{0}\varepsilon\frac{\partial^{2}}{\partial
t^{2}} -  \frac{\partial^{2}}{\partial z^{2}}.
\end{equation}

We choose relation between $\mathcal{A}^{p}_{ln}$ and
$\mathcal{F}^{p}_{ln}$ in form
\begin{equation}\label{EQ:CoffAiF}
\partial_{t}\mathcal{F}^{p}_{ln}=\pm i\partial_{z}\mathcal{A}^{p}_{ln},
\end{equation}
this yield to the two orthogonal polarization which can be write
as
\begin{subequations} \label{E:EFieldCoeff_AiF}
\begin{eqnarray}
E_{r}^{\pm}(r,\varphi,z,t)&=&\mp\frac{1}{2}\sum_{l,n}\frac{1}{\alpha_{ln}}\left[\partial_{z}\mathcal{A}^{\pm}_{ln}J_{l\pm 1}(\alpha_{ln}r) \right]e^{il\varphi}e^{i\omega t-ikz}+c.c.,\\
E_{\varphi}^{\pm}(r,\varphi,z,t)&=&\frac{1}{2}\sum_{l,n}\frac{i}{\alpha_{ln}}\left[\partial_{z}\mathcal{A}^{\pm}_{ln}J_{l\pm
1}(\alpha_{ln}r)\right]e^{il\varphi}e^{i\omega t-ikz}++c.c..
\end{eqnarray}
\end{subequations}
In this case equations \eqref{E:EFieldCoeff_AiF} have simple form
in cartesian co-ordinate system
\begin{subequations} \label{E:EFieldCoeff_AiF_Kart}
\begin{eqnarray}
E_{x}^{\pm}(x,y,z,t)&=&\mp\frac{1}{2}\sum_{l,n}\frac{1}{\alpha_{ln}} \partial_{z}\mathcal{A}^{\pm}_{ln}J_{l\pm 1}(\alpha_{ln}r) e^{i(l\pm 1)\varphi}e^{i\omega t-ikz}+c.c.,\\
E_{y}^{\pm}(x,y,z,t)&=&\frac{1}{2}\sum_{l,n}\frac{i}{\alpha_{ln}}\partial_{z}\mathcal{A}^{\pm}_{ln}J_{l\pm
1}(\alpha_{ln}r)e^{i(l\pm 1) \varphi}e^{i\omega t-ikz}+c.c.,
\end{eqnarray}
\end{subequations}
and we can use it to calculate $P_{z}$.

We take into computation electric field in form
\begin{subequations}\label{EQ:BirifAxis}
\begin{eqnarray}
E_{x}=\frac{1}{2}E_{x}^{+}+\frac{1}{2}E_{x}^{-},\\
E_{y}=\frac{1}{2}E_{y}^{+}+\frac{1}{2}E_{y}^{-}.
\end{eqnarray}
\end{subequations}
For more generality considering about birefringent axis see
\cite{refMenyuk1}.

Let us construct the only transversal mode with fixed $\alpha$ and
$\beta$ which mean that we chose the simplest form by fixing $l=0$
and $n=1$ (we cut series to one term, this allow us to simplify
calculation of the $P_{z}$). We introduce a slowly varying
amplitude of the wave envelope \cite{Hasegawa:book:solitons} in
form
\begin{equation}\label{E:SlowAmplitudeDef}
  \sigma X^{\pm}(\tau,\xi)e^{-ikz},
\end{equation}
where
\begin{subequations}\label{E:SlowCoo}
\begin{eqnarray}
  \xi &=& \sigma z, \\
  \tau &=& (t-k'z)\epsilon,
\end{eqnarray}
\end{subequations}
 where $\sigma$ is nonlinearity parameter and $\epsilon$ is
dispersion parameter.

We insert equations \eqref{E:EFieldCoeff_AiF_Kart} into
\eqref{E:SlowAmplitudeDef} and then we put the slowly varying
amplitude \eqref{EQ:Equ_z_nonlin} in it and if the relation
between parameters is $\epsilon^{2} \sim \sigma$ and if we use a
new coordinate system which moves at group velocity
\eqref{E:SlowCoo}, than to the second order in $\epsilon$ we can
obtain nonlinear Schr\"{o}dinger equation in the form (we don't
get complex conjugate part)
\begin{multline}\label{EQ:NLS_FIN_2P}
\left(i\partial_{\xi}X^{+}+\frac{\epsilon^{2}k''}{2\sigma}\partial_{\tau\tau}X^{+}+
i\partial_{\xi}X^{-}+\frac{\epsilon^{2}k''}{2\sigma}\partial_{\tau\tau}X^{-}\right)e^{i(\omega
t-kz)}= \\=\frac{\varepsilon_{0}\mu_{0}}{\sigma^{2}\pi
N_{01}k}\int\limits_{0}^{r_{0}}\int\limits_{0}^{2\pi}
rJ_{0}(\alpha_{01} r)\frac{\partial^2}{\partial
t^2}P_{z}(X^{+},X^{-})d\varphi dr.
\end{multline}

 Let us now evaluate a right site of equations
\eqref{EQ:NLS_FIN_2P}. If we keep terms up to the order of the
third power of $\epsilon$ and  additional save expression with
$\partial_{\tau}X^{\pm}$ term, we have
\begin{subequations}\label{E:CNLSE}
\begin{eqnarray}
i\partial_{\xi}X^{+}-ik'^{+}_{N}\partial_{\tau}X^{+}+\frac{\epsilon^{2}k''}{2\sigma}
\partial_{\tau\tau}X^{+}
+\mathbb{P}\left[|X^{+}|^{2}+\mathbb{Q}|X^{-}|^{2}\right]X^{+}&=&0,\\
i\partial_{\xi}X^{-}-ik'^{-}_{N}\partial_{\tau}X^{-}+\frac{\epsilon^{2}k''}{2\sigma}
\partial_{\tau\tau}X^{-}
+\mathbb{P}\left[|X^{-}|^{2}+\mathbb{Q}|X^{+}|^{2}\right]X^{-}&=&0,
\end{eqnarray}
\end{subequations}
where

\begin{eqnarray} \label{E:CNLSE_coeff}
\mathbb{P}&=&\frac{3\omega^{2}\mu_{0}\varepsilon_{0}\chi_{xxxx}\sigma}{8
N_{01}k} \int\limits_{0}^{r_{0}}
r\left(\frac{1}{4}J_{0}^{4}(\alpha_{01}r)+\frac{1}{3\alpha_{01}^{2}}J_{1}^{2}(\alpha_{01}r)J_{0}^{2}(\alpha_{01}r)k
^{2}\right)  dr,
\end{eqnarray}
where XPM coefficient (in this case self phase modulation (SPM)
equal $1$)
\begin{equation}\label{EQ:XPMCoeff}
\mathbb{Q}=\frac{1}{\mathbb{P}}\frac{3\omega^{2}\mu_{0}\varepsilon_{0}\chi_{xxxx}\sigma}{8
N_{01}k} \int\limits_{0}^{r_{0}}
r\left(\frac{2}{4}J_{0}^{4}(\alpha_{01}r)\right)  dr,
\end{equation}
and $V^{\pm}_{gN}$ depends on amplitude $X^{\pm}$.
\section{Numerical calculation}\label{SEC:Numerical}
\begin{figure}[htbp]
\centering
\includegraphics[width=0.75\textwidth]{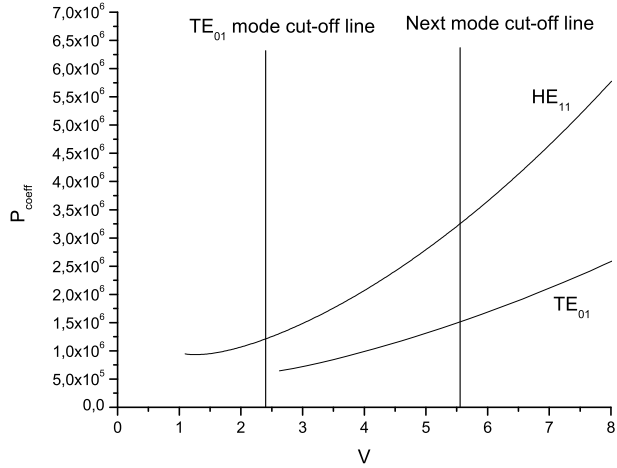}
\caption{Numerical results for $P_{coeff}$ for parameters
\eqref{NumParam}. Vertical line shows mode cut-off. }
\label{PIC:Numsol}
\end{figure}

\begin{figure}[htbp]
\centering
\includegraphics[width=0.7\textwidth]{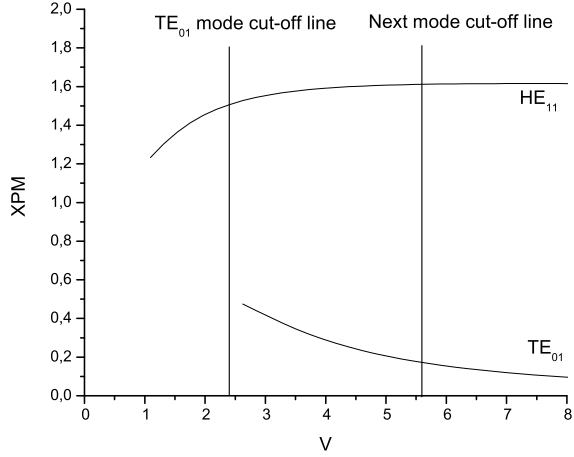}
\caption{Numerical results for XPM coefficient
\eqref{EQ:XPMCoeff}} \label{PIC:numsol_XPM}
\end{figure}

First we define normalized frequency as
\begin{equation}
V=\frac{\omega}{c}r_{0}\sqrt{\varepsilon_1-\varepsilon_2}
\end{equation}
and coefficient which we calculate
\begin{equation}
\mathbb{P}=\chi_{xxxx}\sigma P_{coeff}
\end{equation}
Choosing a value for physical parameters
\begin{subequations}\label{NumParam}
\begin{eqnarray}
    \omega &=& 12.2*10^{14} \quad \text{Hz}\quad (\lambda\approx 1.54\mathrm{\mu m} )\\
    \varepsilon_1&=&2.25 \quad\text{(ref. index $1.5$)}\\
    \varepsilon_2&=&1.96 \quad\text{(ref. index $1.4$)}\\
    \varepsilon_0&=&8.85*10^{-12}\quad \mathrm{\frac{F}{m}}\\
    \mu_0&=&12.56*10^{-7} \quad\mathrm{\frac{N}{A^2}}\\
    r_0&& \text{from $1.2*10^{-6}$m to $10*10^{-6}$m}
\end{eqnarray}
\end{subequations}
On picture \ref{PIC:Numsol} we show results for mode $l=0$ and
$n=0$ (known as $\mathrm{TE_{01}}$) also we show results for mode
$l=\pm 1$ and $n=1$ (known as $\mathrm{HE_{11}}$) which can be
calculate using above procedure (shows here for $l=0$ and $n=0$).
Picture \ref{PIC:numsol_XPM} show XPM coefficient.


\begin{figure}[htbp]
\centering
\includegraphics[width=0.7\textwidth]{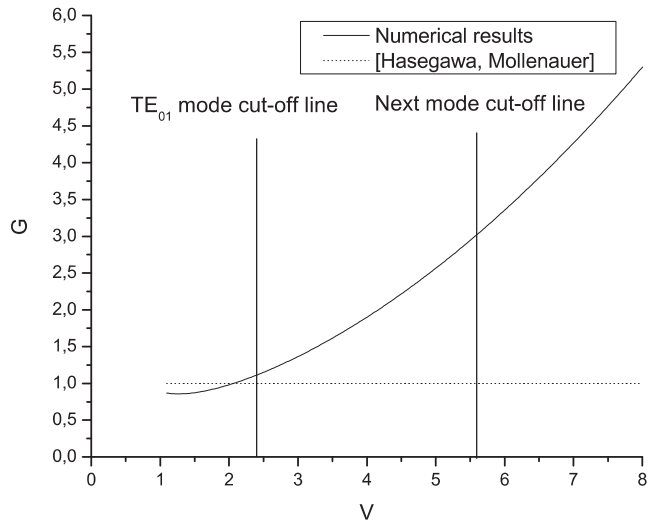}
\caption{Comparison between results} \label{PIC:Compare}
\end{figure}

If we consider only one mode with one polarization we can write
Nonlinear Schr\"{o}dinger Equation (NLS)
\begin{equation}
i\partial_{z}U+\frac{k''}{2}\partial_{tt}U+\gamma|U|^{2}X=0,
\end{equation}
where $\gamma$ is nonlinear coefficient.  We compare $\gamma$ with
work \cite{HasegawaKodama1981,Hasegawa:book:solitons,Mollenauer}
where it was defined as
\begin{equation}\label{Coeff_HasMoll}
\frac{g\omega n_{2}}{c},
\end{equation}
where $g$ depend on variation of the electric field in the fiber
cross section and in most cases takes a value of approximation
$1/2$ and $n_{2}$ is defined by \cite{Hasegawa:book:solitons}
\begin{equation}
n_{2}=\frac{3}{4n}\chi_{xxxx},
\end{equation}
In our case $\gamma$ is define by (for $l=\pm 1$ and $n=1$ with
one polarization)
\begin{eqnarray}
\gamma&=&\frac{3\omega^{2}\chi_{xxxx}}{16 N_{11}k_{11}c^{2}}
\int\limits_{0}^{r_{0}}
r\left[\frac{1}{2}J_{1}^{4}(\alpha_{11}r)+\frac{k_{11}
^{2}}{3\alpha_{11}^{2}}J_{1}^{2}(\alpha_{11}r)\left(J_{0}^{2}(\alpha_{11}r)+J_{2}^{2}(\alpha_{11}r)\right)\right]dr,
\end{eqnarray}
We don't have dependence on mode cross-section (radius of light
beam and fiber) but it is included by the boundary condition. On
picture \ref{PIC:Compare} we make comparison our results to
\eqref{Coeff_HasMoll}. On graph $G$ is defined as
\begin{equation}
G = \frac{\gamma(\text{our numerical results})c}{g\omega n_{2}}
\end{equation}

\section{Conclusion}
In this paper we shows a new approach to derive a formula for CNLS
equations.

The main idea is to take into account multi-mode case. Here we
show the simplest case for $l=0,\pm 1$ and $n=1$ (as single mode)
but if we use equation \eqref{EQ:Equ_z_nonlin} and take
electromagnetic field \eqref{E:EFieldCoeff_AiF_Kart} with more
modes we could compute multi-mode case. The proceeding is the same
as for single but it is more intricate and derive simple formula
is more difficult (because we have additional terms corresponding
to mode interaction).

In this paper we don't show formula for $V^{\pm}_{gN}$ but it can
be calculate.

Additionally we can allow  for change birefringent axis (see eq.
\eqref{EQ:BirifAxis}) and take case with different grup velocity
(introduce $k_{+}\neq k_{-}$).

The definition of the nonlinear coefficient is used in works
relating to quantum effects in fibers \cite{VVKozlov}, and  the
coefficient is defined likely in the non-quantum NLS equation.
\section{Acknowledgements}
The second author would like to thank K.J.Blow, V.Mezentsev and
S.Turisin for useful advices and discussion during a stay at Aston
University grant also we would like to thank Y.Kodama for valuable
counsel.

The work is supported by the Polish Ministry of Scientific
Research and information Technology grant PBZ-Min-008/P03/2003.

\end{document}